\documentclass[a4paper,11pt]{article}
\usepackage{graphicx,amsmath,amssymb}

\usepackage{makeidx}
\usepackage{graphicx}
\usepackage{subfigure}

\begin{document}

\title{Is the pinning of ordinary dislocations in $\gamma$-TiAl intrinsic or
  extrinsic in nature? A combined atomistic and kinetic Monte Carlo approach}

    \author{I. H. Katzarov and A. T. Paxton}
 
      \date{}

     \maketitle
 
 \begin{center}

   \textit{ Atomistic Simulation Centre, School of Mathematics and Physics,  
 Queen's University Belfast, Belfast BT7 1NN, Northern Ireland, U.K.}

  \end{center}

\begin{abstract} 

We address the question of the observed pinning of
$\frac{1}{2}\left\langle 1\bar 10 \right]$ ordinary screw
dislocations in $\gamma$-TiAl which leads to the characteristic
trailing of dipoles in the microstructure. While it has been
proposed that these may be variously intrinsic or extrinsic in
nature, we are able to rule out the former mechanism. We do
this by means of very large scale, three dimensional atomistic
simulations using the quantum mechanical bond order
potential. We find that the kink-pair formation energy is
large---6~eV, while the single kink migration energy is
conversely very small---0.13~eV. Using these, and other
atomistically derived data, we make kinetic Monte Carlo
simulations at realistic time and length scales to simulate
dislocation mobility as a function of stress and temperature.
In the temperature range of the stress anomaly in
$\gamma$-TiAl, we determine whether one or several of the pinning
and unzipping processes associated with generation of jogs are
observed during our simulations. We conclude that the pinning of ordinary
dislocations and anomalous mechanical behaviour in
$\gamma$-TiAl must be attributed to a combination of extrinsic obstacles
 and extensive cross-slip in a crystal containing impurities.
  
\end{abstract}
  
\section{Introduction}

It is well known that the deformation behaviour of $\gamma$-TiAl
is complex due to the tetragonality of the $L1_{0}$ structure.
The $\gamma$ phase of TiAl can deform by several deformation
modes: glide of ordinary dislocations, glide of superlattice
dislocations and twinning~\cite{{App}}.  The ${\bf b}=\frac{1}{2}\left\langle
1\bar 10 \right]$ ordinary dislocations in deformed TiAl
alloys exhibit a unique morphology consisting of numerous
pinning points along the dislocation line aligned roughly along
the screw dislocation direction, and bowed-out segments between
the pinning points~\cite{{App}}.  Post mortem TEM analyses
\cite{{Gre},{Vig1},{Sri}} have shown that the screw segments
separated by pinning points are lying in different parallel
$(111)$ planes as a result of a cross-slip mechanism. Extensive
cross-slip was also observed in {\it in situ} straining experiments
\cite{{Cou1}}.  Yield stress anomaly (increase of the yield
stress with temperature) has been observed in $\gamma$-TiAl
compounds and ascribed to the $\frac{1}{2}\left\langle 1\bar 10
\right\rangle\{111\}$ slip system
\cite{{Vig1},{Inu},{Nak}}.  There is compelling microscopic
evidence that pinning plays a prominent role in the anomalous
mechanical behaviour.  The determination of the subsequent
cusp unpinning mechanism is of fundamental importance in the
interpretation of the stress anomaly.

Deep cusps and the trailing of dipoles suggest that the
dislocations have many jogs.  At the microscopic scale, the
pinning points have been analysed as jogs aligned along the screw
direction, whose density increases with temperature in the domain
of the stress anomaly~\cite{{Lou1},{Lou2}}.  The exact nature of
the obstacles developed on these cross-kinks seems to be
controversial.  Viguier {\it et al.}~\cite{{Vig1}} and Sriram
{\it et al.}~\cite{{Sri}} have concluded that the pinning points
are formed as a result of intrinsic processes involving single or
double cross-slip mechanisms leading to formation of jogs.  The
three dimensional structure formed during collision of kinks
during their lateral propagation cannot move in the direction of
the dislocation glide.  Explanations based on the
activation of cross-slip suffer however from not explaining
satisfactorily a number of experimental results:

\begin{enumerate}

\item Non-screw ordinary dislocations, Shockley dislocations
  involved in twinning and superlattice dislocations are also
  anchored at many points~\cite{{Mes},{Hau},{Gre2},{Cou2}}.

\item Ordinary dislocations moving in the same plane are able to
  anchor at exactly the same location in the sample~\cite{{Cou1}}.

\item An annealing treatment which precipitates interstitial
  atoms in excess decreases the density of cusps and debris~\cite{{Gre}}.

\end{enumerate}

To explain these experimental results, it was concluded that
pinning points are extrinsic in nature and due to some chemical
heterogeneity~\cite{{Gre},{Hau},{Zgh}}.  Messerschmidt {\it et
al.}~\cite{{Mes}} and Morris~\cite{{Mor}} have ascribed these
pinning points to small extrinsic obstacles like oxygen atoms or
Al$_{2}$O$_{3}$ precipitates.

Understanding the glide mechanism of ordinary dislocations in
different stress and temperature ranges can help us to determine
which of its main features is likely to account for the strength
anomaly of $\gamma$-TiAl.  In addition, the behaviour of these
ordinary dislocations deserves particular attention since they
are responsible for most of the deformation in two-phase lamellar
TiAl alloys~\cite{{Kat1},{Kat2},{Kat3},{Kat4}}.

The present work aims at exploring the mobility of
$\frac{1}{2}\left\langle 1\bar 10 \right]$ ordinary screw dislocations
in $\gamma$-TiAl through simulation of the specific mechanisms
of motion of an individual dislocation in the $L1_{0}$ structure.

Due to the high Peierls barrier~\cite{{Kat1},{Por}}, the motion
of ordinary screw dislocations in TiAl is believed to be
controlled by nucleation, migration and annihilation of kinks.
When the stress on a dislocation is lower than its Peierls
stress, the dislocation line stays at rest in a given lattice
position, interrupted by the thermally assisted process of kink
pair nucleation.  The separation of kinks under the influence of
stress and thermal activation results in translation of the
entire dislocation line to the next lattice position. The overall
dislocation movement is a cumulative effect of a large number of
individual kink events.

Dislocation dynamics (DD) simulations~\cite{{Bul},{Tan}} employ
simple rules that dictate the motions of dislocations which are
represented as interconnected small straight segments in an
elastic continuum connected through nodes. Local stresses
resulting from applied loading and internal stresses are computed
on each of those segments.  The dynamics of dislocation motion is
thus reduced to the dynamics of the nodes prescribed by the
mobility rules.  Such a description replaces the true kink
dynamics of dislocation glide with a mean field measure of the
average mobility of a dislocation line.  The reliability with
which the DD simulations mimic the plastic behaviour of the
crystal depends on the accuracy of the mobility rules.

In order to overcome these limitations and to obtain a more
realistic description of the dynamics of a dislocation line over
long time scales we carry out kinetic Monte Carlo (kMC) simulations
of the mobility of $\frac{1}{2}\left\langle 1\bar 10 \right]$
dislocations.  Several researchers have performed kMC studies
of dislocation behaviour to provide a more realistic link
between kink dynamics and the averaged dynamics of a
dislocation line~\cite{{Cai1},{Cai2},{Deo}}.  The advantage of this
approach is that it replaces arbitrary assumptions about the
nature of dislocation mobility with input based upon
microscopic understanding. In addition, kMC can achieve
extended time scales for simulation of the dislocation
mobility.  kMC simulations use atomistic results for the nature
of the core energetics, short ranged dislocation segment
interactions, and rate theory to simulate dislocation mobility
as a function of stress and temperature.

In our simulations we assume that the ordinary dislocation moves
by the kink mechanism through adjacent Peierls valleys.  We
examine ordinary dislocation mobility in TiAl at fixed stresses
and temperatures based upon atomistic input.  The rates of kink
pair nucleation and kink migration events are expressed in terms
of their respective energy barriers, which are computed using
atomistic simulation techniques.  The kMC model allows
dislocations to cross-slip onto secondary glide planes. Thus,
motion of screw dislocations becomes three dimensional and can
involve a number of unusual mechanisms affecting dislocation
mobility.  In the temperature range of the stress anomaly, we
check whether one or several of the above mentioned pinning and
unzipping processes associated with generation of jogs are
observed during simulations.

\section{kMC model}
\label{sec_kMC}

Ordinary screw dislocations in TiAl do not dissociate into a
planar structure due to the high energy of the complex stacking
fault (CSF) in $\gamma$-TiAl~\cite{{Por}}.  Instead, the
non-planar core of these dislocations spreads symmetrically on
two cross-slip $\{111\}$ planes (Fig.~1) and their glide is not
confined to a single glide plane. Cross-slip should readily occur
by which screw dislocations change their glide planes. In fact,
each screw dislocation segment can nucleate kink-pairs on either
of the two $\{111\}$ planes.

The kMC model does not consider any details of the core
structure. It focuses on dislocation motion on length and time
scales far greater than those of atomistic simulations. The key
idea of kMC is to treat dislocation motion as a stochastic
sequence of discrete rare events whose mechanisms and rates are
computed within the framework of transition state theory.
Dislocations are represented as interconnected small straight
segments in an elastic continuum.  Local stresses resulting from
applied loading and internal stresses are computed on each of
those segments.

In the present model, we study the ordinary screw dislocation
represented by a piece-wise straight line stretched along the
$[1\bar 10]$ direction.  While the dislocation has, on average, a
screw orientation, it consists of screw (S) and edge (E)
dislocation segments such that kinks on the screw dislocation are
perfect edge segments.  E-segments all have the same length $h$,
the unit kink height, while S-segments can be of any length. In
the present work we choose for the length of the smallest
S-segment $a=b$, where ${\bf b}=\frac{1}{2}\left\langle 1\bar 10 \right]$
is the Burgers vector of the screw dislocation.  The time and
3D space is discretised in the form of a square grid.  Here,
$a$ and $h$ are the grid spacings in the screw and edge
directions, respectively.

The screw dislocation in TiAl was allowed to move on the two
$\{111\}$ glide planes that comprise the
$\frac{1}{2}\left\langle 1\bar 10 \right\rangle\{111\}$ slip system.  Kink
pairs are allowed to nucleate on any part of S-segments and in
any of the two $\{111\}$ glide planes intersecting the $[1\bar 10]$
direction. Once nucleated, a kink (E-segment) can move in its
glide plane along the dislocation line until it recombines with
another kink with the opposite sign. Periodic boundary
conditions are applied so that kinks leaving at one end
re-enter the dislocation from the other end.  During kMC
simulations, the dislocation moves under the action of external
stress through kink-pair nucleation, migration and
recombination. This model does not allow for climb of the edge
segments.

The kMC simulation samples different classes of possible kink
nucleation and migration events.  Kink\textendash kink
annihilation is considered as a special case of kink migration.
The kMC methods use information on the energetic barriers to
atomic hopping or kink migration that control the motion of a
dislocation in order to obtain the relationships between the
velocity of the dislocation and the driving forces acting upon
it.  The kinetics of dislocation motion is completely specified
by the matrix of transition rates.  Transition-state theory
expresses the rates of kink-pair nucleation and kink migration
events in terms of their respective energy barriers, which can be
computed using atomistic simulations.  The following form of
transition-rate matrix element provides a physically consistent
description of the energetics and thermally activated formation
of kink-pairs~\cite{{Hir}}:

\begin{equation*}
 J_{\hbox{\scriptsize k-p}}(n)=\omega\>\exp\left[ -\frac{\Delta
     E_{n}-TS-\sigma_{\hbox{\scriptsize eff}}\>b\Delta A(n)/2}{k_{\rm B}T} \right]  
\end{equation*} 

$ \Delta E_{n} $ is the formation energy of a kink-pair,
$k_{\hbox{\scriptsize B}}$ is the Boltzmann constant and $\omega$ is
the pre-exponential “frequency” factor, which is set equal to the
Debye frequency.  $\sigma_{\hbox{\scriptsize eff}}$ is the local
resolved stress that includes the applied stress and the
self-stress field.  The self stress is the stress exerted on a
dislocation segment by all other dislocation segments and can be
calculated from the gradient of the total energy of the
dislocation configuration due to virtual displacements of the
segment~\cite{{Hir}}.  The total energy is computed by using
non-singular continuum theory of dislocations \cite{{WC}}. The
major advantage of this theory is that it contains no
singularities.  The singularity intrinsic to the classical
continuum theory is removed by spreading the Burgers vector
isotropically about every point on the dislocation line using a
spreading function characterized by a single parameter $r_{c}$,
the spreading radius.  The non-singular expression for
dislocation energy depends on the choice of the cut-off parameter
$r_{c}$.  We determined it by comparing the energy of a kink-pair
predicted by non-singular continuum theory with atomistic
data. $S$~is usually taken to be $3k_{\hbox{\scriptsize B}}$ \cite{{Hir}}.  The
last term is the work done by the external applied stress and
$\Delta A(n)$ is the change of the area of the glide plane as a
result of kink-pair formation.

A kink, once formed, can migrate along the dislocation line. The
rate of migration of the kinks depends on the magnitude of the
activation energy for kink motion $W_{n}$ (secondary Peierls
barrier).  If this is very large the kink migration is thermally
activated with transition rate given by

\begin{equation}
\label{eq_rate}
 J_{\hbox{\scriptsize m}}=\omega\>\exp\left[ -\frac{W_{n} - TS -\sigma_{\hbox{\scriptsize eff}}b^{2}h/2}{k_{\hbox{\scriptsize B}}T} \right] 
\end{equation} 

If the secondary Peierls barrier for the kink motion is very
small the kink migration is not thermally activated but is
controlled by phonon drag. In such a case, the kink velocity
$v_{\hbox{\scriptsize k}}$ is proportional to the driving force experienced by the
kink.  In the kMC model the thermally activated migration events
specified by the transition rates (\ref{eq_rate}) are replaced by continuous
movement of the kinks with velocity proportional to the stress of
the dislocation line calculated at this moment
\cite{{Deo},{Cai4}}.

\subsection{Sustainable kink-pair nucleation}

The dislocations in the kMC model are discretised as sets of
screw and edge segments.  Our atomistic simulations, described in
the following section, predict that a screw segment of the
dislocation line moves to an adjacent Peierls valley upon
nucleation of kink-pair only if the distance between the kinks is
greater than the kink width $w_{c}=16b$.  Only in this case can the
kink-pair be represented by set of screw and edge segments
with length $h$.  If a kink-pair is formed with a width smaller
than a critical width $w_{c}$ the height of the kinks is smaller
than the grid spacing in the edge direction. In this case the
kink-pair can not be represented by a continuous piecewise
straight line on the discretised space. In addition, the elastic
energy of such configurations can not be calculated accurately by
using non-singular continuum theory of dislocations.  A solution
to this problem lies in nucleating a kink-pair only of finite
width $w_{in}$ which is greater than the critical width
$w_{c}$. Such a kink-pair, once formed, can be represented by a
continuous piecewise straight line on the discretised space and
its elastic energy can be calculated correctly.  To be able to do
that, it is necessary to compute the rate at which such kink
pairs should be nucleated. We determine the rate for the
formation of a kink-pair of width $w > w_{c}$ by using survival
probabilities analysis on the 1D random walk. Details of this
method can be found elsewhere~\cite{{Deo2}}.

The survival probability $p(1\rightarrow n)$, which is defined as
the probability that a kink-pair of width $w=1b$ expands to
$w=nb$ before it shrinks to $w=0$, can be obtained from
\begin{equation*}
 p(1\rightarrow n)=\prod_{k=1}^{n}p(k\rightarrow k+1)
\end{equation*} 
and $p(k\rightarrow k+1)$ can be calculated from the following recursive formulae
\begin{equation*}
 p(k\rightarrow k+1)=\frac{F(k)}{1-[1-F(k)]\>p(k-1 \rightarrow k)}
\end{equation*} 
$F(k)=J^{+}(k)/J^{-}(k)$ is the probability that upon escape from state $kb$ the kink-pair will expand to width $(k+1)b$. 
$J^{+}(k)$ and $J^{-}(k)$ are the rates of of forward and backward jumps from state $k$ as follows:
\begin{align}
 J^{+}(k)&=\omega\>\exp\left(-\frac{[E(k+1)-E(k)]-TS-\sigma_{\hbox{\scriptsize eff}}\>b\Delta
   A(k)/2}{k_{\hbox{\scriptsize B}}T}\right)[1+H(k)] \label{eq_rates+} \\
 J^{-}(k)&=\omega\>\exp\left(-\frac{[E(k-1)-E(k)]-TS-\sigma_{\hbox{\scriptsize eff}}\>b\Delta A(k)/2}{k_{\hbox{\scriptsize B}}T}\right)[1+H(k-1)] \label{eq_rates-}
\end{align} 
where $H(k)=1$ if $k \geq 1$ and $H(k)=0$ if $k \leq 0$.  With
this definition the rate of embryonic kink-pair nucleation is
simply $J(1)=J^{+}(0)$.  $E(k)$ is the energy of two kinks
separated by a distance $kb$.  The approximation we employ here
by describing the rates of forward and backward jumps in the
forms~(\ref{eq_rates+}) and~(\ref{eq_rates-}) is that the local
stress remains constant over the time it takes for an embryonic
kink-pair to expand to with $nb$. This approximation is fairly
good because, under the conditions considered here, the expansion
time is usually very small compared to the time over which the
dislocation configuration and the local stresses change
significantly.  When the distance between the kinks $kb$ is
smaller than the critical width $w_{c}$, $E(k)$ is specified by
the kink-pair formation energy which is calculated atomistically
(section~\ref{sec_kink-pair}, Fig.~4, below).  The rate at which
kink-pairs of width $w_{in}b$ greater than the critical width
$w_{c}$ appear can be obtained from the rate of embryonic
kink-pair nucleation $J(1)$ and the corresponding survival
probability

\begin{equation}
\label{eq_J}
      J(n)=J(1)\>p(1\rightarrow n)
\end{equation} 
                     
In our kMC model only kink-pairs with width $w_{in}>w_{c}$ are
allowed to nucleate on S-segments with appropriate length
$L_{S}>w_{c}$.  The nucleation rate of such kink-pairs is given
by~(\ref{eq_J}).

\section{Atomistic simulations}

A crucial input for kMC simulations is the information from
atomistic simulations of the properties of dislocation cores and
short range dislocation segment\textendash segment interactions,
such as the energies of a single isolated and kink-pair and
the associated Peierls stress for kink motion.  As both the
qualitative and quantitative outcome of the kMC simulations is
extremely sensitive to these input quantities, the accuracy of
the atomistic simulations is of utmost importance.

The atomistic simulation of individual dislocations requires
special treatment due to the long ranged elastic field associated
with them.  For long straight dislocations, we may use a
periodicity along the dislocation line, but should avoid using
periodic boundary conditions in the other two directions since
the dislocation strain field is very long ranged. The cells used
in these simulations, as in most real space calculations, are
divided into two regions of relaxing and non relaxing atoms, the
latter bounded by a free surface sufficiently far from the core
to avoid image stresses.  

Pettifor's bond order potential (BOP)~\cite{{Pet},{Aok}}
represents a numerically efficient scheme that works within the
orthogonal tight binding approximation with environment dependent
matrix elements. The multi-atom character of the forces, is
thereby captured in a physical way that goes beyond the standard
pair functionals (Finnis--Sinclair or embedded atom
method).  Apart from their genuine quantum-mechanical origin,
BOPs have two additional important advantages. First, the
evaluation of the energy scales linearly in computational time
with the number of atoms, and, second, the real-space formalism
avoids the need of imposing full periodic boundary conditions
common to k-space methods. Both of these features are crucial for
studies of dislocations since such simulations often require a
large number of atoms and complex geometries.  The atomistic
simulations of the the properties of ordinary screw
dislocations in this study have been made using the BOP for
$\gamma$-TiAl, constructed and extensively tested against
accurate first-principles methods~\cite{{Zna}}.

Molecular statics (MS) techniques, which seek to minimise the
total energy of the system at zero-temperature, can be used for
relaxation of the inner atomistic regions of the simulation cell,
where the initial atomic positions in the computational cell are
established by the conditions of linear anisotropic elasticity.
Traditionally, fixed boundary conditions have been most often
used in such dislocation simulations.  This requires very large
simulation cells in practice, but this method is always
problematic with respect to force build-up between fixed and
relaxed atomic regions.  The size of the atomistic region can be
dramatically reduced by using so-called flexible boundary
conditions.  An elastic Green's function version of such
conditions have been developed for both 2D and 3D dislocation
simulations, known as Green's function boundary conditions
(GFBC)~\cite{{Sin},{Rao}}.  The GF techniques allow dislocation
configurations to be simulated atomistically with very little
accumulation of forces at the boundaries of the simulation cells.
In this method, a buffer layer is introduced between the fixed
outer and inner relaxed atomistic regions of the simulation cell,
allowing one to dynamically update the boundary conditions of the
simulation, while dramatically reducing the size of the atomistic
region.

Using the BOP method for the description of the interatomic
interactions, we have implemented GFBC to calculate the
properties of ordinary screw dislocations through MS
simulations. These properties include the core structure and
energy, the kink-pair formation energy, the primary and
secondary Peierls stresses as well as the kink motion
activation energies.  In this way the simulated 3D dislocation
configurations can be contained in significantly reduced
atomistic region without compromising the fidelity of the final
core structure and energy. 

\subsection{Isolated kink}

Two 2D core structures of an infinite $\frac{1}{2}\left[ 1\bar 10
  \right] $ screw dislocation were used to obtain the initial
approximation to the structure of a kink.  These relaxed
configurations of the 2D ordinary screw dislocation were
constructed by using two different elastic centres in an initial
approximation to the dislocation displacement field.  Anisotropic
elasticity theory was used to introduce a $\frac{1}{2}\left[
  1\bar{1}0 \right] $ screw dislocation in the simulation cell. A
cylindrical crystal, one periodic unit in length along the
dislocation line direction, $[1\bar{1}0]$, was constructed in
order to obtain the two symmetry-related core structures of the
ordinary screw dislocation.  The $x$ and $y$ axes in the
simulation cell were $[11\bar{2}]$ and $[111]$ respectively.  The
centres of the two core structures were introduced at two
adjacent Peierls valleys with a relative displacement of $h$, the
periodic length along the $[11\bar{2}]$ direction.  Atomistic
relaxation was used to optimise the atomic positions in the
active region. The 2D GF procedure was applied to relieve the
boundary forces.  One of the relaxed configurations obtained for
the core structure of straight $\frac{1}{2}\left[ 1\bar{1}0
  \right] $ screw dislocations is shown in a differential
displacement plot in Fig. 1.

The two 2D cores are then used to obtain the initial
approximation to the structure of an isolated kink on an ordinary
screw dislocation.  A 3D cylindrical cell is constructed with a
total length of $45b$, where $b$ is the Burgers vector of the
dislocation.  The two halves of the simulation cell were made up
of 22 periodic units of the two relaxed configurations of the 2D
$\frac{1}{2}\left[ 1\bar{1}0 \right] $ screw dislocation
respectively. The central cell of unit length $b$ was defined as
the average of the positions of identical atoms in the two 2D
core structures~\cite{{Rao}}.  This overall length $45b$ is
chosen to allow a smooth transition from one core structure to
the next.  Using a simple continuum estimate of the width of the
kink~\cite{{Rao},{Petu}} we obtain a kink width of $13b$.
Therefore $45b$ is a reasonable choice for the overall length of
the simulation cell.

Relaxation is carried out in the inner atomistic region and the
GF technique is used to relax the forces that develop in the
buffer GF layer. The core structure of the central cell of the
kinked dislocation line was characterised using differential
displacement plots (Fig. 2).  The difference between the energies
of the kinked and straight dislocation gives the single-kink
formation energy.  The formation energy of an isolated kink in
$\gamma$-TiAl calculated by our atomistic simulations is 2.88~eV.

In order to determine the width of the kink we plot the
differences $\Delta z =[z_{3}+z_{4}-(z_{1}+z_{2})]/2$ between the
average $z$-coordinates of the atoms designated numbers 1,~2 and
3,~4 in Fig. 2. This plot allows us to illustrate the position of
the dislocation lines corresponding to the initial approximation
of the isolated kink and its relaxed core structure (Fig. 3).
The width of the single-kink in TiAl predicted by our atomistic
simulations is $16b$. This value is in good agreement with the
continuum estimate of the width of the kink.

\subsection{Kink-pair}
\label{sec_kink-pair}

The energy of two sharp kinks as a function of their separation
$L$ can be determined quite accurately when they are well
separated~\cite{{Hir}}.  The kink-pair can be regarded as being
composed of straight line segments and the energy is calculated
as the energy of the interactions between the line segments,
including the self-interaction. The energy of two kinks can be
separated into individual kink energies $2E_{\hbox{\scriptsize k}}$ and interaction
energy $E_{i}$.

\begin{equation*}
  E(L)=2E_{\hbox{\scriptsize k}}+E_{i}
\end{equation*} 
                
The energy of two mixed orientation, ``oblique'' kinks is
expressed equivalently provided that kink-pair is separated by a
distance $L$ greater than the kink width $w$.  When this
condition does not hold $E(L)$ can no longer be separated into
individual kink energies and interaction energy.  The continuum
estimation of the elastic energies of such configurations is
difficult.

Applying the BOP-GFBC technique we determine the core structure
and energy of a kink-pair as a function of the distance between
the kinks.  The two relaxed configurations of the 2D screw
dislocation were used to obtain initial approximations to the
structure of two pure edge kinks separated by distances from $1b$ to
$23b$.  Core structures were relaxed to obtain the energy of the
kink-pairs.  The results are plotted in Fig.~4.  The differences
$\Delta z$ between the average $z$-coordinates of atoms $1, 2$
and $3, 4$ which illustrate the shape of the dislocation lines
for several kink-pairs are shown in Fig.~5. The atomistic
simulations predict that a screw segment of the dislocation line
between both kinks transfers to the adjacent Peierls valley only
if the distance between the kinks $L$ is greater than the kink
width, ({\it i.~e.} $L>16b$).  The formation energy of two kinks
in this case can be determined quite accurately by using
non-singular continuum theory of dislocations~\cite {{Hir}}.  By
comparing the energies of two kinks separated by distances
$L>16b$, predicted by non-singular continuum theory, with the
corresponding atomistic data we can determine the spreading
radius (see section~\ref{sec_kMC}) or cut-off
parameter. The value of the cut-off parameter obtained in
this way is $r_{c}=0.125b$.

We made continuum calculations of the formation energy both for
pure edge, or ``perfect'', and oblique kinks. In both cases we
obtained similar results for the energy of the kinks. The
relative discrepancy between the formation energies of perfect and
oblique isolated kinks is 0.25. For kink-pairs separated by
distance $L> w$ we find that the relative discrepancies between
the energies of perfect and oblique kinks is about 0.5.  Taking
into account these calculations, we approximate the real oblique
kinks on ordinary screw dislocation in our kMC model by perfect
edge segments.  The approximation of perfect kinks is appropriate
when the kink width is smaller than the distance between two
kinks in a kink-pair.  In this case, calculation within the
non-singular continuum theory and perfect kinks approximation
provides a correct result for the self-stress exerted on a
dislocation segment.

\subsection{Kink migration}

In order to estimate the energy barrier for kink migration we
started with a simulation block containing the fully relaxed core
structure of an isolated kink on an ordinary screw dislocation.
To this we applied an incrementally increasing pure shear stress
such that the force on the dislocation pushes the kink along the
dislocation line.  In practice this stress is imposed by applying
the appropriate homogeneous shear strain, which is evaluated with
anisotropic elasticity theory. This strain was superimposed on
the dislocation displacement field for all the atoms. Relaxation
was carried out at every incremental step in the applied stress,
until the kink migrates one crystal period along the dislocation
line.  At each step we calculated the change in the total energy
of the cell as the kink migrates from its initial to final
position.  In particular, the maximum energy difference along the
path is the energy barrier of migration of a single kink.  Fig.~6
shows the energy of an isolated kink during its motion
between two adjacent Peierls valleys. The energy is plotted at
the nine incrementally increasing shear stresses that we have
been working with.  We obtain 0.13~eV for the energy of
migration of a single kink on a screw ordinary dislocation in
$\gamma$-TiAl.

In contrast to high Peierls barrier for kink-pair nucleation
($\sim$6~eV, see Fig.~4), kink motion on the ordinary screw
dislocation in TiAl should be very easy, as manifested in the low
kink (or secondary) Peierls barrier. A high mobility of nucleated
kinks should be expected given that the applied stress in our
simulations (rather less than the Peierls stress
$\sigma_{P}=0.04C_{44}$\cite{{Por}}) far exceeds the secondary
Peierls stress ($0.0023C_{44}$) determined in our atomistic
simulations.  The low energy barrier and secondary Peierls stress
mean that kink migration along the ordinary screw dislocation is
not thermally activated but is limited by phonon drag.
Therefore, the kink velocity in the present kMC simulations is
calculated as

\begin{equation*}
        v_{\hbox{\scriptsize k}}=\sigma_{g}\,\frac{b}{B}
\end{equation*}
where $\sigma_{g}$ is the glide component of the resolved shear
stress, $b$ the Burgers vector and $B$ the phonon drag
coefficient.  In our simulations we use the same numerical value
of the drag coefficient for kink motion, $B = 1.5\times
10^{-5}$~Pa~s, as in Fivel {\it et al.}~\cite{{Fiv}}.

\section{Results}

In this section we present the results of kMC simulations of
ordinary screw dislocation motion under $0.75\sigma_{P}$ shear
stress with different orientations of the maximal resolved shear
stress (MRSS) plane. The total length of the S-segments are
$4000b \simeq 1.1~\mu$m while each E-segment is only
6.5~\AA. Since the typical measured distance separating two
pinning points is 0.1--0.15~$\mu$m~\cite{{Cou2}}, the choice of
this length of the dislocation line in our simulations is to
maintain enough space for intrinsic pinning of the ordinary
dislocation.  Two different shear stress orientations of the MRSS
plane have been tested in our kMC simulations: ({\it i\/}) single
slip conditions, and ({\it ii\/}) non symmetrical double slip
conditions, each of them at different temperatures between 600
and~1200~K.  The choice of such a moderately high stress and
temperature values in these simulations is to maintain a
sufficient number of kinks on the dislocation. In the case of
single slip conditions the MRSS plane is parallel to the primary
glide plane $(111)$.  The resolved shear stress on the primary
glide plane is much larger than that in the secondary plane, and
kink-pairs mostly nucleate on this plane.  If the pinning
points are supposed to be intrinsic, they are initiated by the
collision of superkinks nucleated in both primary and cross-slip
planes, and transformed into jogs by double cross-slip.  In the
case of non symmetrical double slip conditions the MRSS plane
bisects two glide planes $(111)$ and $(1\bar{1}1)$, making
kink-pair nucleation almost equally probable on both planes. We
expect that collision of kinks nucleated in both primary
and cross-slip planes resulting in self-pinning should be readily
observed in this case.

However, for both shear stress orientations of the MRSS plane the
simulation results were very surprising, in that there is always
only one kink-pair along the entire dislocation line.  Kink
nucleation is rare and existing kinks tend to glide along the
dislocation line through the periodic boundary conditions and
recombine before the next kink nucleation event occurs.  The
averaged kink-pair nucleation rate per lattice site on the
glide plane is $J_{\hbox{\scriptsize k-p}} = 1.5\times 10^{5}$~$b^{-1}$~s$^{-1}$, and
the average kink velocity is $v_{\hbox{\scriptsize k}} \approx 10^{14} b$~s$^{-1}$.
The interplay between the rate of kink-pair nucleation and the
kink migration velocity gives rise to a particular length scale
$L_{c}$ such that when a dislocation is shorter than $L_{c}$,
there is always only one kink-pair along the entire dislocation
line.  For typical kink-pair nucleation rate and kink migration
velocity in our simulations we find that this length scale is more
than $10 \mu$~m. Comparison with the typical observed distance
separating two pinning points indicates that the probability for
self-pinning and an intrinsic nature of the pinning points of screw
ordinary dislocation in $\gamma$-TiAl is negligible.

In order to demonstrate self-pinning of the ordinary dislocation
we artificially reduced the kink migration velocity by increasing
the phonon drag coefficient. With decreasing kink migration
velocity the number of kink-pairs nucleated on both glide planes
along the dislocation line steadily increases.  At $B=5 \times
10^{-3}$~Pa~s, corresponding to a kink migration velocity $v_{\hbox{\scriptsize k}}
\approx 3\times 10^{11} b$~s$^{-1}$ a few pinning points are
generated along the dislocation line, leading to cusp formation,
see Fig.~7.  The rate of self-pinning and cusp formation along
the screw direction is higher in the case of non symmetrical
double slip conditions.  The process of intrinsic pinning
is initiated by collision of two kink-pairs formed in both primary
and cross-slip planes.  Such elementary jogs grow in size when
more kinks pile-up on either sides of the initial pinning point
forming superjogs. The kinks in the pile-ups on two sides of the
pinning point belong to different parallel primary glide planes.
If the dislocation lines on two sides of the pinning point cross
each other they may reconnect by recombination of kink-pairs. As
a result, the dislocation line is now reduced in size, leaving
behind a prismatic loop as shown in Fig.~7.  The cusps are easily
released by this unzipping mechanism and straight dislocations
along the screw direction are restored.

The separation between pinning point and the unzipping are
functions of temperature and shear stress orientations of the
MRSS plane.  In the case of non symmetrical double slip
conditions pinning, unzipping and formation of prismatic loops
aligned along the screw direction is frequently observed in our
kMC simulations.  At lower temperatures and single slip
conditions the rate of self-pinning and unzipping is less. Due
to the lower unzipping rate, in this case we observed formation
of dipoles as a result of propagation of screw dislocation line
on both sides of the pinning points.

\section{Discussion}

The results of our simulations indicate that in the unit
processes of ordinary dislocation glide described by the kink
pair formation and migration model, nucleation of sustainable
kink-pairs is rate limiting, and hence plays an important role in
determining both the velocity of the dislocation and its glide
mechanism.

The specific ordinary dislocation mechanisms expected in the
$L1_{0}$ structure were explored by Fivel~{\it et
  al.}~\cite{{Fiv}} through an adaptation of a 3D mesoscopic
dislocation simulation initially developed by Kubin {\it et
  al.}~\cite{{Kub}}.  Using simple rules for kink-pair nucleation
and propagation on ordinary screw dislocations, the simulation
reproduced the main features of the ``local pinning unzipping''
(LPU) mechanism~\cite{{Lou3}} (pinning on cross-slip generated
jogs, cusp unzipping, formation of aligned prismatic loops). It
was concluded that extrinsic obstacles do not seem to be
necessary to account for both the observed microstructures nor
the stress anomaly~\cite{{Fiv}}. This conclusion concerning the
dislocation glide mechanism was a result of the the high
kink-pair nucleation rate employed by Fivel~{\it et
  al.}~\cite{{Fiv}}. The high probability of thermally activated
kink-pair nucleation followed from low kink-pair activation
energy, 0.45~eV, which they used.  Since no experimental
measurement of the activation energy is available owing to the
stress anomaly, the activation energy in~\cite{{Fiv}} is simply
taken to be in the range of the analytically calculated
activation enthalpy for the cross-slipping back from the
secondary $(1\bar{1}1)$ cross-slip plane~\cite{{Lou2}}.

On the other hand, our atomistic simulations show that the
kink-pair activation energy is one order of magnitude higher,
which in its turn leads to a much lower thermally activated
kink-pair nucleation rate.  The interplay between the rate of
kink-pair nucleation and the high kink migration velocity due to
low secondary Peierls barrier leads to low density of kink-pairs
in both $\{111\}$ glide planes at length scales at which pinning
of the dislocation line should be expected.  In the absence of
obstacles, kinks on the ordinary screw dislocation have very high
mobilities, limited predominantly by lattice drag resistance. In
the presence of solute atoms, the rate of kink migration is
determined by the solute--kink interaction.  The
solute--dislocation interaction can be divided into two parts
depending on the distance of the solute from the dislocation
segment; one associated with the long ranged elastic fields of the
solute and the other with the short ranged solute–-core
interaction.  If the solute is close to the dislocation core, the
core--solute interaction provides a barrier $E_{b}$ to the motion
of the kink. Usually kinks require thermal activation to overcome
the local energy barrier $E_{b}$ associated with solute atoms
near the core. In this case, the rate of kink migration is
determined by the magnitude of the solute--kink interaction energy
$E_{b}$.

Owing to high kink-pair activation energy and high kink mobility
in the absence of interaction with extrinsic obstacles predicted
by the present atomistic simulations, the probability for
self-pinning is negligible, and therefore the nature of the
pinning points on screw ordinary dislocations in pure $L1_{0}$
crystal cannot be intrinsic.  It follows that the observed
dislocation pinning must be attributable to extrinsic
obstacles. These obstacles presumably reduce kink migration
velocity, thus increasing the density of kink-pairs in both
$\{111\}$ glide planes at length scales at which pinning of the
dislocation line should be expected.  The increased density of
kink-pairs formed in both primary and cross-slip planes leads to
higher probability for collision of two kinks and formation of
jogs.

The results of our simulations indicate that the pinning on screw
ordinary dislocations can be attributed to the relation between
extrinsic obstacles and extensive cross-slip.  As shown in the
present study, the lateral propagation of kinks occurs very
quickly and this implies a very small probability of movements in
the adjacent slip plane of the screw segments which are lying in
the sessile configuration before this lateral propagation.  If a
part of a dislocation moving without any interaction with
extrinsic obstacles makes a double cross-slip, the whole
dislocation moves onto a primary glide plane parallel to the
initial one since no obstacles oppose the lateral propagation of
the edge segments lying in the cross-slip plane.  If the
dislocation moves in a crystal containing extrinsic obstacles,
the number of kink-pairs nucleated on both primary and cross-slip
planes along the dislocation line steadily increases as a result
of the reduced mobility of kinks due to the solute--dislocation
interaction.  The increased density of kink-pairs nucleated in
both glide planes leads to higher probability for collision of
two kinks and formation of jogs by double cross-slip. The
pinning points and dipoles dragged by screw dislocations are thus
probably formed through this combination of extrinsic obstacles
and cross-slip in a crystal containing impurities.

\section{Conclusions}

In our kMC simulations of ordinary dislocation mobility in TiAl
we examined an isolated dislocation moving at applied stress by
the kink propagation mechanism through adjacent Peierls valleys
in the $L1_{0}$ crystal structure of $\gamma$-TiAl.  The rates of
kink nucleation and migration events were expressed in terms of
their respective energy barriers, which were computed using fully
three dimensional atomistic simulations employing Green's
function boundary conditions and a quantum mechanical
prescription for the interatomic forces.  The results of the
present simulations indicate that the interplay between the rate
of kink-pair nucleation and the kink migration velocity leads to
a surprisingly low density of kink-pairs in both glide planes at
length scales at which intrinsic pinning of the dislocation line
should be expected.  Therefore, the probability of intrinsic
pinning, which is initiated by the collision of kinks nucleated
in both primary and cross-slip planes, and transformed into jogs
by double cross-slip is negligible. 

If we artificially reduce the kink migration velocity the number
of kink-pairs nucleated on both glide planes along the
dislocation line steadily increases.  The increased density of
kink-pairs formed in both primary and cross-slip planes leads to
higher probability for collision of two kinks and formation of
jogs, demonstrating that our method is capable of producing the
intrinsic self-pinning, albeit at unphysical values of the
parameters.

{\it We conclude that the pinning points are not intrinsic in nature.}
The pinning of ordinary dislocations and anomalous mechanical
behaviour in $\gamma$-TiAl must therefore be correlated to extrinsic
obstacles.  Amongst the other possible mechanisms, obstacles
could reduce kink migration velocity, thus increasing the density
of kink-pairs in both glide planes at length scales at which
pinning of the dislocation line should be expected.

\section*{Acknowledgements}

This research was supported by the EPSRC, under Grant No. EP/E025854/1

\newpage
 
 Fig. 1. Differential displacement plot showing the screw
 components of a $\frac{1}{2}\left[1\bar{1}0 \right]$
 screw dislocation. The arrows joining atom pairs indicate the
 relative displacements of the two atoms in the direction normal
 to the page; hence the arrows show the screw component of the
 relative displacement. The length of the arrow (in arbitrary
 units) is proportional to the amount of relative displacement.

 Fig. 2. Differential displacement plot showing the core
 structure of the central cell of the kinked dislocation
 line. The arrows are as in Fig.~1. This shows the plane of atoms
 perpendicular to the Burgers vector in the region separating two
 long screw dislocations having elastic centres displaced by the
 kink height, $h$. Hence a kink is lying along the $[11\bar{2}]$
 direction in the plane of the diagram. The core of the two
 $\frac{1}{2}\left[1\bar{1}0 \right]$ screw dislocations
 spreading onto two adjacent $(111)$ and two adjacent $(1\bar
 11)$ planes can thereby be clearly seen.
 
 Fig. 3. Core spreading of an isolated kink. The relative
 displacement, $\Delta z =[z_{3}+z_{4}-(z_{1}+z_{2})]/2$ between
 the average $z$-coordinates of the atoms designated numbers 1,~2
 and 3,~4 in Fig.~2 is plotted against distance from the kink
 centre in units of $b$. In the ideal, unrelaxed case this
 amounts to a ``perfect'' kink. The relaxed structure represents
 what in the text we refer to as an ``oblique'' kink, or non
 pure edge kink. We see that the width of a relaxed kink is on
 the order of ten Burgers vectors.

 Fig. 4. Kink-pair formation energy as a function of the
 separation between the kinks. Here we constrain the kink-pair
 separation in the 3D simulation as described in the text and
 show the energy of the corresponding relaxed kink-pair.

 Fig. 5. The shape of the dislocation lines for kink-pairs with
 different distances $w$ between the kinks. As in Fig.~3, we show
 the relative displacement, $\Delta z$, this time at the cores of
 a kink-pair, constrained as in Fig.~4 to a separation $w/b$,
 this being indicated on each set of data. The dotted line at
 $w=21b$ refers to an ``ideal'', unrelaxed kink-pair. This
 illustrates the point we make in the text that for a kink-pair
 to be identifiable in a kMC simulation its separation must be
 greater than $\sim 16b$, so that it extend into the next Peierls
 valley. 
 
 Fig. 6. The  energy of an isolated kink during transition
 between two adjacent secondary Peierls valleys. The applied
 stress is in units of the shear elastic constant, $c_{44}$.

 Fig. 7. A typical outcome of the kMC simulations, showing the
 cusped character of the dislocation line observed at reduced
 kink mobility. The MRSS plane bisects two glide planes $(111)$ and
 $(1\bar{1}1)$ making kink-pair nucleation almost equally probable
 on both planes. The process of self-pinning is initiated by 
 the collision of kinks nucleated in both primary and cross-slip planes.
 The cusps are released by recombination of the  kink-pairs on
 both sides of a pinning point, leaving behind debris (prismatic loops).

\newpage
\includegraphics[]{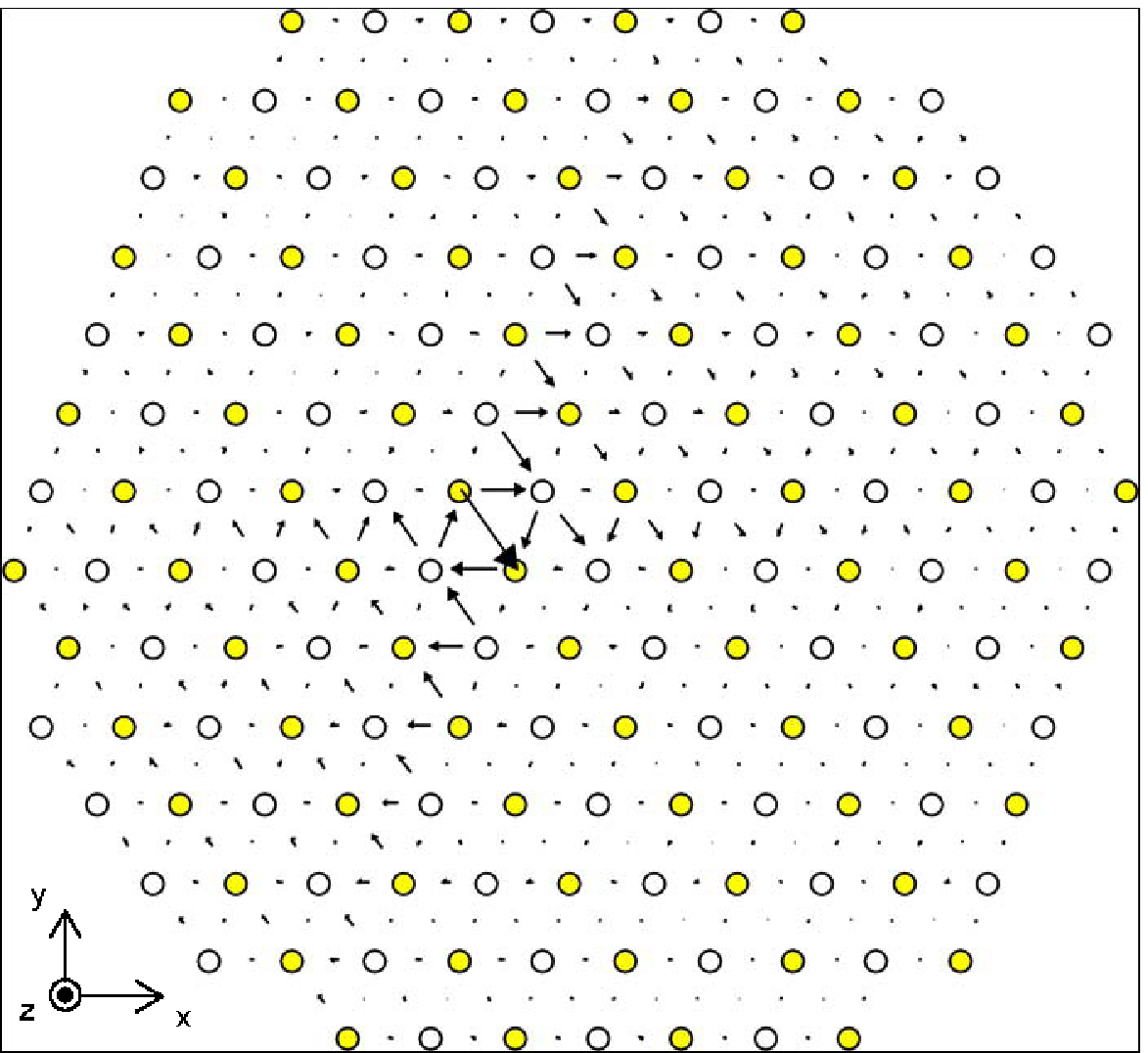}

\vskip 1cm
\centerline{Fig. 1}

\newpage
\includegraphics[]{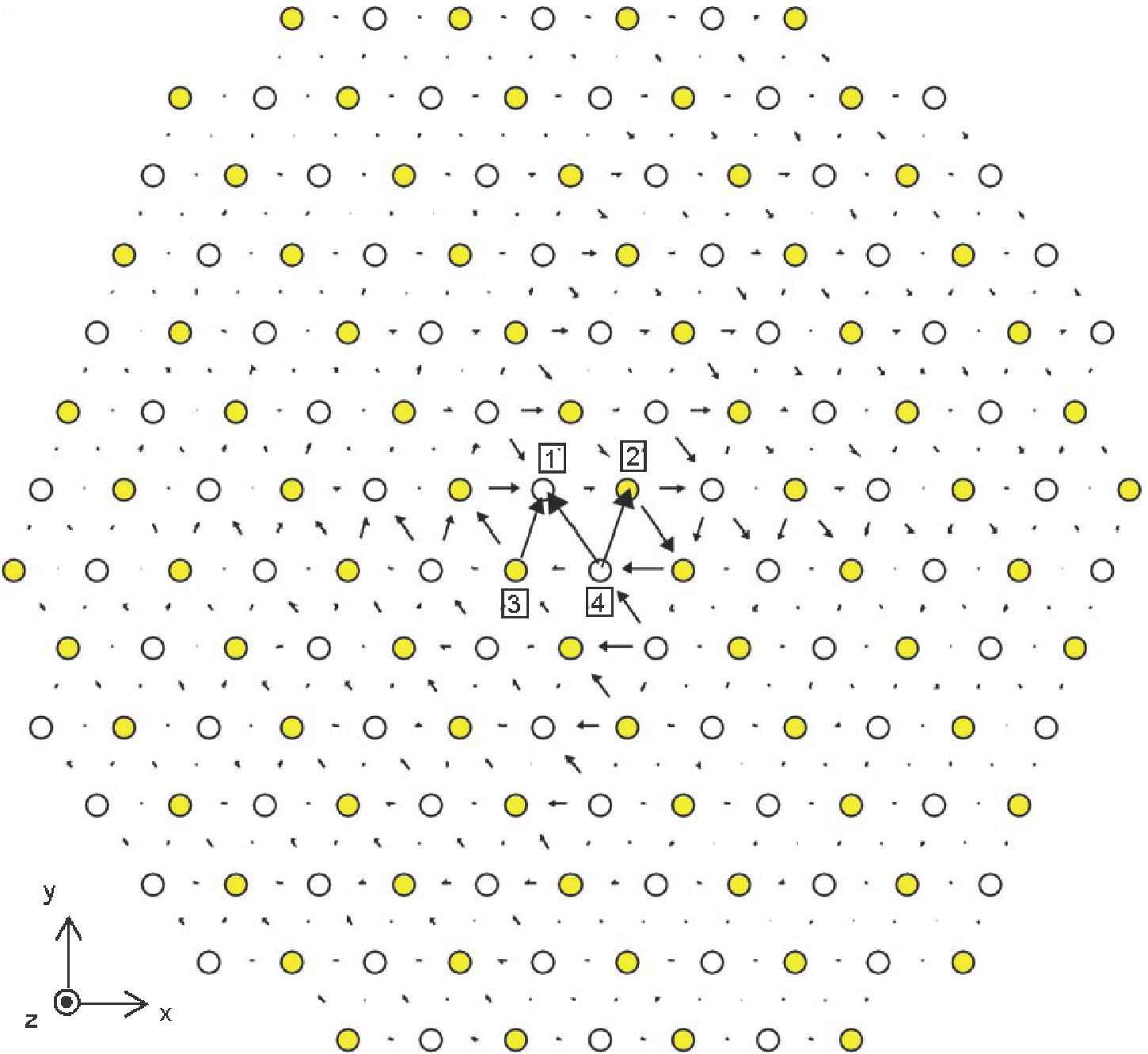}

\vskip 1cm
\centerline{Fig. 2}

\newpage
\includegraphics[scale=0.8]{fig3-redraw.ps}

\vskip 1cm
\centerline{Fig. 3}

\newpage
\includegraphics[scale=0.8]{fig4-redraw.ps}

\vskip 1cm
\centerline{Fig. 4}

\newpage
\includegraphics[scale=0.8]{fig5-redraw.ps}

\vskip 1cm
\centerline{Fig. 5}

\newpage
\includegraphics[scale=0.8]{fig6-redraw.ps}

\vskip 1cm
\centerline{Fig. 6}

\newpage
\hskip -2.5cm
\includegraphics[scale=0.6]{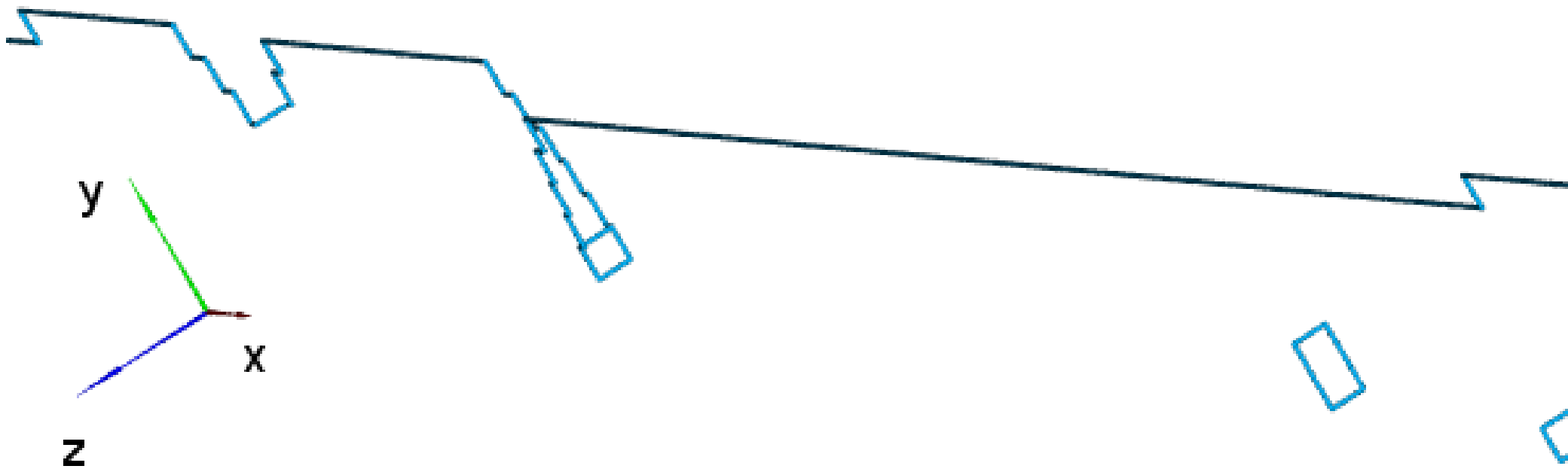}

\vskip 1cm
\centerline{Fig. 7}


\begin{thebibliography}{1}


\bibitem{App} Appel, F., Wagner, R., Mater. Sci. Eng. 1998; R22:187-268.
\bibitem{Gre} Gregori F, 1999, PhD thesis, Universite de Paris VI. 
\bibitem{Vig1} Viguier B, Hemker KJ, Bonneville J, Louchet F, Martin JL. Phil. Mag. A, 1995;71:1295-312.
\bibitem{Sri} Sriram S, Dimiduk DM, Hazzledine P, Vasudevan VK. Phil. Mag. A, 1997;76:965-93.
\bibitem{Cou1} Couret A., Phil. Mag. A, 1999;79:1977-94.
\bibitem{Inu} Inui H, Matsumuro M, Wu DH, Yamaguchi M., Phil. Mag. A, 1997;75:395-423.
\bibitem{Nak} Nakano T., Hagihara K., Seno T., Sumida N., Yamamoto M., Umakoshi Y., Phil. Mag. Letters, 1998;78:385-91.
\bibitem{Lou1} Louchet F., and B. Viguier, Phil. Mag. A, 1995;71:1313-33.
\bibitem{Lou2} Louchet F. and B. Viguier, Phil. Mag. A, 2000;80:765-79.
\bibitem{Mes} Messerschmidt U., Bartsch M., Haussler D., Aindow M., Hattenhauer R., Jones I.P..
 Mater. Res. Soc. Symp. Proc., 1995;364:47-52.
\bibitem{Hau} Haussler D., Bartsch M., Jones I.P., Messerschmidt U. Phil. Mag. A, 1999;79:1045-71.
\bibitem{Gre2} Gregori F. and Veyss\`{e}re P., Phil. Mag. A, 2000;80:2913-33.
\bibitem{Cou2} Couret A., Intermetallics, 2001;9:899-906.
\bibitem{Zgh} Zghal S., Menand A., Couret A., Acta Metall. Mater., 1998;46:5899-905.
\bibitem{Mor} Morris M.A., Intermetallics, 1996;4:417-26.
\bibitem{Kat1} Katzarov I. H., Cawkwell M. J., Paxton, A. T. and Finnis, M. W., Phil. Mag, 2007;87:1795-1809.
\bibitem{Kat2} Katzarov I. H. and Paxton, A. T., Phil. Mag, 2009;89:1731-50.
\bibitem{Kat3} Katzarov I. H. and Paxton, A. T., Acta Mat., 2009;57:3349-66.
\bibitem{Kat4} Katzarov I. H. and Paxton, A. T., Phys. Rev. Letters, 2010;104:225502.
\bibitem{Por} Porizek R., Znam S., Nguyen-Manh D., Vitek V. and Pettifor D.G.,
 Mat. Res. Soc. Symp. Proc., 2003;753:BB4.3.1-6.
\bibitem{Bul} Bulatov, V., et al., Nature, 1998;301:669-72.
\bibitem{Tan} Tang M., L.P. Kubin, G.R. Canova, Acta Mater., 1998;9:3221-35.
\bibitem{Cai1} W. Cai, V.V. Bulatov, J.F. Justo, S. Yip, A.S. Argon, Phys. Rev. Lett., 2000;84:3346-49. 
\bibitem{Cai2} W. Cai, V.V. Bulatov, S. Yip, J. Comput. Aided Mater. Design, 1999;6:175-183.
\bibitem{Deo} Deo C.S., D.J. Srolovitz, W. Cai, V.V. Bulatov, J. Mech. and Phys. of solids, 2005;53:1223–47.
\bibitem{Cai4} W Cai, V.V. Bulatov, S. Yip, A.S. Argon, Mater. Sci. and Eng. A, 2001;309-310:270-73.
\bibitem{Hir} Hirth, J. P. and Lothe, J., Theory of Dislocations, 2nd edn. Wiley-Interscience, New York, 1982.
\bibitem{WC} W. Cai, A. Arsenlis, C. Weinberger, V. Bulatov, J. Mech. Phys. Solids, 2006;54:561-87.
\bibitem{Deo2} Deo C.S. and D. J. Srolovitz, Phys. Rev B, 2001;63:165411.; V.V.Bulatov and Wei Cai, 
Computer Simulations of Dislocations, Oxford University Press, 2006.
\bibitem{Pet} Pettifor, D. G., Phys. Rev. Lett., 1989;63:2480-3.
\bibitem{Aok} Pettifor D. G., Aoki M. Philos. Trans. Roy. Soc. London Ser A – Math Phys Eng Sci 1991;334:439-49.
\bibitem{Zna} Znam S., Nguyen-Manh D., Pettifor D. G.  and Vitek V., Phil. Mag A, 2003;83:415-38.
\bibitem{Sin} Sinclair, J. E., J. appl. Phys., 1971;42:5321-9.
\bibitem{Rao} Rao S. , C. Hernandez, J. P. Simmons, T. A. Parthasarathy and C. Woodward, Phil. Mag. A, 1998;77:231-56.
\bibitem{Petu}Petukhov , B. V., Phys. Metals Metallogr. (USSR), 1083;56:123-9.
\bibitem{Fiv} Fivel M.C., F. Louchet, B. Viguier and M. Verdier, Mat. Res. Soc. Symp. Proc., 2001;646:N7.10.1-10.6.
\bibitem{Kub} Kubin L.P., G.R. Canova, M. Condat, B. Devincre, V. Pontikis and Y. Br\'{e}chet, in Solid State Phenom.,
1992;23/24: 455-72.
\bibitem{Lou3} Louchet, F., and Viguier B., Scripta metall., 1994;31:369-74. 
 




\end{thebibliography}
\end{document}